\documentclass{article}
\usepackage{spconf,amsmath,graphicx,verbatim}

\newcommand{\OMIT}[1]{}

\usepackage{multirow}
\usepackage{booktabs}
\usepackage{tabularx}
\usepackage{array}
\usepackage{floatrow}
\usepackage{csquotes}
\usepackage[table,xcdraw]{xcolor}

\usepackage{mathtools}

\usepackage{hyperref}
\newcolumntype{Z}{>{\centering\let\newline\\\arraybackslash\hspace{0pt}}X}

\makeatletter
\def\bstctlcite{\@ifnextchar[{\@bstctlcite}{\@bstctlcite[@auxout]}}
\def\@bstctlcite[#1]#2{\@bsphack
  \@for\@citeb:=#2\do{%
    \edef\@citeb{\expandafter\@firstofone\@citeb}%
    \if@filesw\immediate\write\csname #1\endcsname{\string\citation{\@citeb}}\fi}%
  \@esphack}
\makeatother

\newcommand\blfootnote[1]{%
  \begingroup
  \renewcommand\thefootnote{}\footnote{#1}%
  \addtocounter{footnote}{-1}%
  \endgroup
}

\title{Voice Filter: Few-shot Text-To-Speech speaker adaptation using Voice Conversion as a post-processing module}

\name{Adam Gabryś$^{\star}$, Goeric Huybrechts$^{\star}$, Manuel Sam Ribeiro$^{\star}$, Chung-Ming Chien$^{\dagger}$, Julian Roth$^{\star}$,}
\secondlinename{Giulia Comini$^{\star}$, Roberto Barra-Chicote$^{\star}$, Bartek Perz$^{\star}$, Jaime Lorenzo-Trueba$^{\star}$}
\address{$^{\star}$ Alexa AI \qquad $^{\dagger}$ National Taiwan University (NTU)}
\begin{document}
\ninept

\maketitle

\begin{abstract}
State-of-the-art text-to-speech (TTS) systems require several hours of recorded speech data to generate high-quality synthetic speech. When using reduced amounts of training data, standard TTS models suffer from speech quality and intelligibility degradations, making training low-resource TTS systems problematic.  In this paper, we propose a novel extremely low-resource TTS method called Voice Filter that uses as little as one minute of speech from a target speaker. It uses voice conversion (VC) as a post-processing module appended to a pre-existing high-quality TTS system and marks a conceptual shift in the existing TTS paradigm, framing the few-shot TTS problem as a VC task. Furthermore, we propose to use a duration-controllable TTS system to create a parallel speech corpus to facilitate the VC task. Results show that the Voice Filter outperforms state-of-the-art few-shot speech synthesis techniques in terms of objective and subjective metrics on one minute of speech on a diverse set of voices, while being competitive against a TTS model built on 30 times more data.\footnote{Audio samples will be made publicly available with a related blog post on  \href{www.amazon.science}{www.amazon.science}.}\blfootnote{\hspace*{-1.3mm}$^{\dagger}$Work carried out as an intern.}
\end{abstract}


\section{Introduction}

State-of-the-art text-to-speech (TTS) technologies are capable of generating high-quality synthetic speech on a variety of situations.
In order to achieve very high quality, TTS typically requires several hours of studio-quality data, drawn from either a single or multiple speakers \cite{latorre2019effect}.
As such, reducing the amount of speech data to a few hours imposes limitations on the quality and intelligibility of those systems \cite{chung2019semi}.
Because it is not always feasible to collect several hours of speech data, particularly when scaling TTS voices to a large number of new speakers, the problem of building low-resource TTS voices has been thoroughly explored \cite{chen19f_interspeech, zhang2020unsupervised, xu2020lrspeech, huybrechts2021low}.
When the aim is building high-quality TTS systems in extremely low-resource scenarios, such as when only one minute of speech from the target speaker is available, we primarily aim to capture speaker identity and we must defer the modelling of phonetic and prosodic variability to supplementary speech data.

Common approaches to this problem therefore rely on speaker adaptation, whereby the parameters of a multi-speaker model are optimized on a few samples from the target speaker \cite{ICASSP2021_VC_Challenge, taigman2017voiceloop, chen2018sample}.
The adaptation process aims to modify only the speaker identity, which are the speech attributes defining the target speaker as an individual. To control speaker identity in few-shot speaker adaptation, there are techniques such as Vector-Quantized models \cite{oord2017neural,razavi2019generating,wu2020vqvc}, U-Net structures \cite{ronneberger2015u, wu2020vqvc}, attention mechanisms \cite{choi2020attentron} or a combination of loss functions \cite{wang2020bi, cai2020speaker}.
Alternatively, speaker identity can be controlled through representations that are inherited from external speaker verification systems or jointly trained with the primary model \cite{arik2018neural,jia2018transfer,cai2020speaker}.
During adaptation, studies have proposed to optimize all model parameters \cite{chen2018sample,kons2019high}, selected components \cite{moss2020boffin,zhang2020adadurian, chen2021adaspeech}, or to focus instead on external speaker representations \cite{chen2018sample,arik2018neural}. An alternative approach is to address the adaptation problem via data augmentation. This can be done through conventional signal processing techniques \cite{lorincz2021speaker}, but there are more sophisticated methods that proposed producing high-quality synthetic data for the target speaker through a voice conversion (VC) model \cite{huybrechts2021low, shah21_ssw}. The TTS system is then optimized from scratch on a mixture of natural and synthetic data.

There are, however, shortcomings associated with these methods.
Wang et al \cite{wang2020spoken} suggests that, with the speaker adaptation strategy, a single architecture becomes responsible for the modelling of linguistic content and speaker identity.
As in this framework it is not entirely clear which model parameters are responsible for speaker identity, the impact of parameter adaptation can be diluted. Additionally, fine-tuning a complex architecture on a few samples can easily lead to model over-fitting, reducing overall quality and intelligibility.
On the other hand, data augmentation-based approaches, still require at least 15 minutes of training data from the target speaker in order to optimise the TTS models. As such they aren't directly applicable to very low-resource scenarios.

In this paper, we address the problem of extremely low-resource TTS by using VC as a post-processing module which we refer to as \enquote{Voice Filter} on top of a high-quality single-speaker TTS model. This single-speaker TTS model is also used to generate a synthetic parallel corpus for the Voice Filter training. Our proposal presents the following novelties and advantages:
(1) The overall process becomes modular, splitting it into speech content generation task followed by a speaker identity generation one. This improves efficiency, robustness and interpretability, as well as enabling task-dependent adaptation.
(2) We leverage the strengths of parallel VC without assuming that a parallel corpus is available by synthetically generating frame-level matching speech pairs via a duration-controllable TTS model. 

In summary, we split the problem of traditional few-shot TTS voice creation into two tasks: speech content and speaker identity generation. The split means we can lower the amount of speech required to train a synthetic voice for a particular speaker identity down to as little as one minute of speech by limiting the complexity of the problem. This translates in the produced synthetic speech quality being comparable to those of TTS models trained on 30 times more data.

\begin{figure*}[!th]
\begin{center}
\includegraphics[scale=0.14]{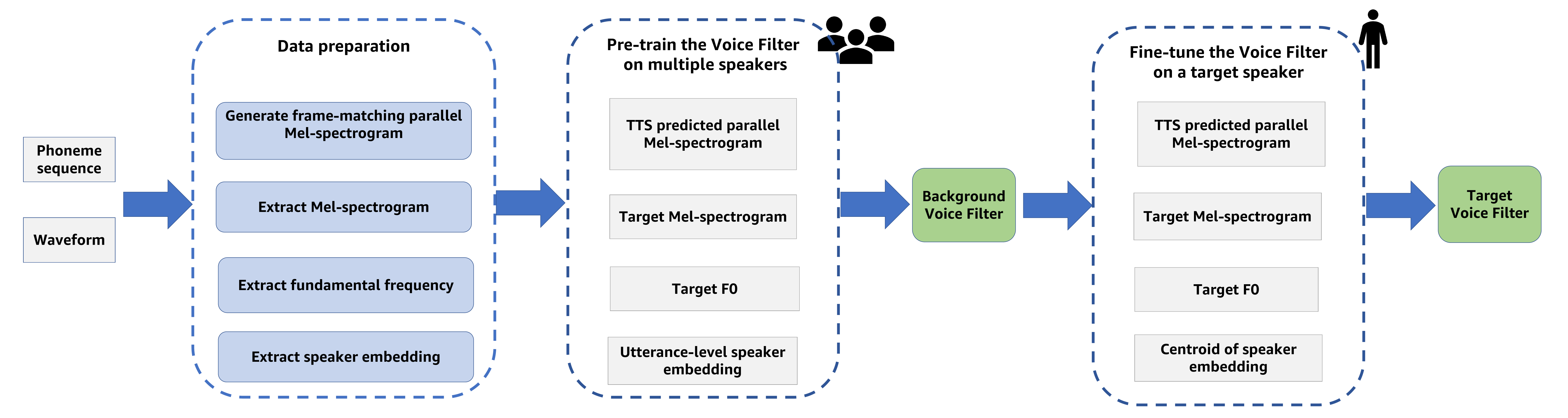}
\caption{Synthetic parallel data preparation and training flowchart for the proposed Voice Filter.}
\label{fig:vf_training_pipeline}
\end{center}
\end{figure*}
\newpage

\section{Method}

Voice Filter approaches the problem of extremely low-resource TTS voice building by decoupling the speech content and the speaker identity generation into two separate tasks, with the Voice Filter itself focusing on the latter. This results in more modularization and more robust speaker identity generation compared to adapting a multi-speaker TTS model. Since the Voice Filter is in charge of speaker identity generation only, it operates at a lower abstraction level (Mel-spectrograms) than the overall TTS system (phonemes). We believe that this speech-to-speech task is easier than the text-to-speech one, especially in very low-resource settings.
  
The proposed VC module (Voice Filter) is located between a single-speaker duration-controllable TTS model \cite{shah21_ssw} and a universal neural vocoder \cite{oord2018parallel}. This enables us to generate Mel-spectrograms for any desired text with the TTS model, which is subsequently given the appropriate speaker identity by the Voice Filter and ultimately converted into a time-domain waveform by the vocoder. 

\subsection{Creation of synthetic parallel corpus}
\label{subsec:parallel_corupus}

The proposed method begins by synthetically generating a parallel dataset to enable training of our Voice Filter (Figure~\ref{fig:vf_training_pipeline}, Data preparation). 
Working with a synthetic parallel corpus allows us to overcome two of the biggest limitations of parallel VC, while still leveraging its strengths: (1) It is no longer required to have a large parallel speech corpus between source and target speakers, which is hard and expensive to collect; and (2) It is no longer necessary to apply duration warping methods to align the parallel corpora across the multiple speakers. This is possible because the input to the Voice Filter is duration-controlled constructed synthetic speech and not recordings. However, this means that in order to generate the parallel corpus we require two different datasets: a single-speaker corpus for the TTS system and a multi-speaker corpus for the Voice Filter training.

Creating the synthetic parallel corpus consists of 3 steps:
\begin{enumerate}
	\item Force-align all available data at the phone level, which we did using the pre-trained Kaldi ASpIRE TDNN system \cite{peddinti2015jhu}, including both the single-speaker and multi-speaker corpora.
	\item Train a duration-controllable TTS system \cite{shah21_ssw}, making use of the single-speaker corpus.
	\item Generate synthetic data matching the transcripts and phone-level durations of the multi-speaker corpus that was aligned in the preliminary step using the trained duration-controllable TTS system.
\end{enumerate}

This three-step process results in a parallel corpus that is aligned at the frame-level between synthetic single-speaker and natural multi-speaker speech samples. It forms the training data for the proposed Voice Filter, which to the best of our knowledge, is a novel inclusion to the VC problem.

For the purpose of this paper and experimentation, the single-speaker corpus contains 10 hours of high-quality speech data read in a neutral speaking style by a male US English speaker.
The multi-speaker corpus consists of 120 gender-balanced male and female US English speakers, with approximately 40 minutes of data per speaker with complete phonetic coverage.

\subsection{Model architecture}
\begin{figure}[!t]
\begin{center}
\includegraphics[scale=0.12]{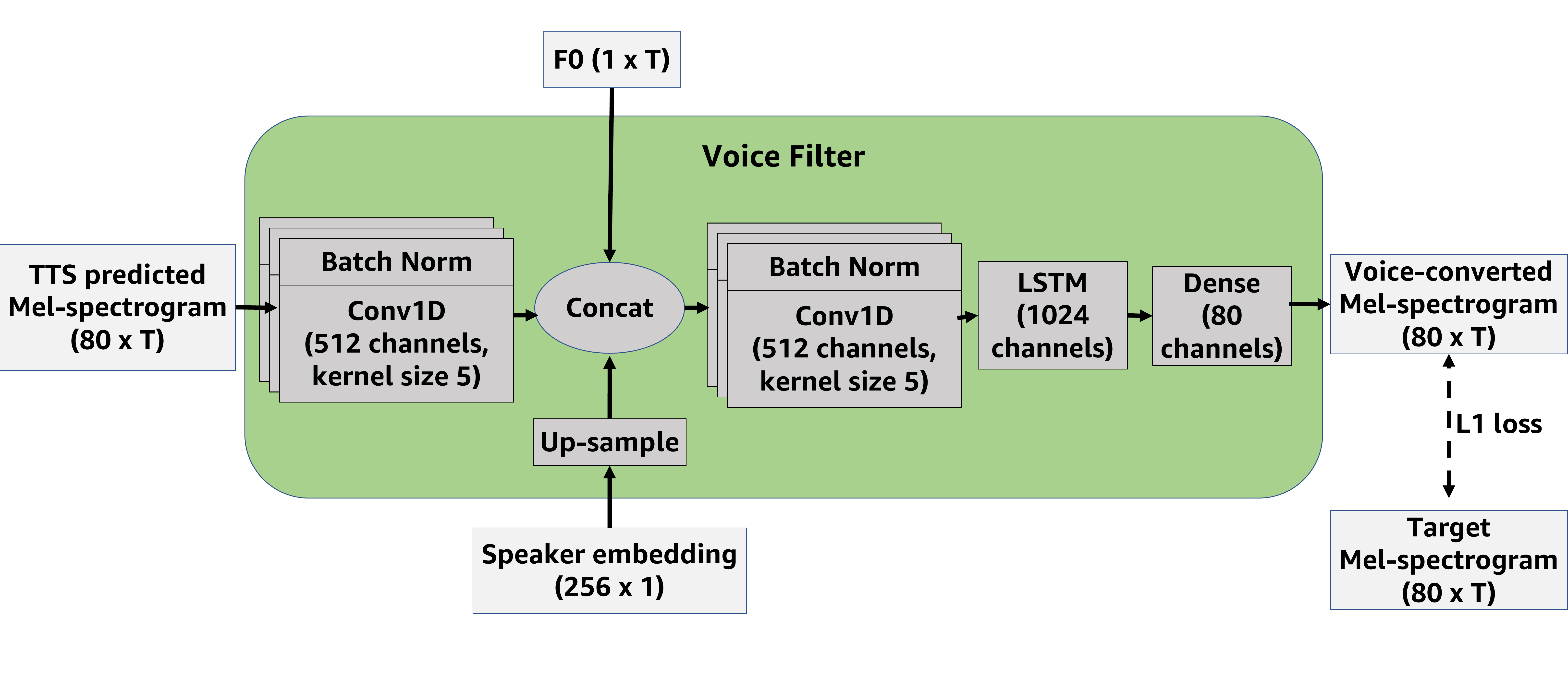}
\caption{Proposed Voice Filter architecture.}
\label{fig:vf_architecture}
\end{center}
\end{figure}

The Voice Filter model (Figure~\ref{fig:vf_architecture}) takes as inputs and outputs 80-bin Mel-spectrograms of equal-length and consists of a 6-stack of size-preserving 1D convolutions with 512 channels and a kernel size of 5 with batch-norm. This is followed by a uni-directional LSTM and a Dense layer with 1024 nodes.

We concatenate the target speaker embedding and log-f0 contour to the hidden representation of the third convolution layer. The speaker embedding is a 256-dimensional vector defined at the utterance level and broadcasted to the frame level. The speaker verification system used to extract the embeddings was trained on the multi-speaker corpus and optimized on a Generalized End-to-End Loss \cite{wan2018generalized}. We observed that the log-f0 contour helps the model better absorb the prosodic differences between the input and target speaker. Effectively speaking, this means that the Voice Filter doesn't need to learn how to adjust prosody information between the source and the target speakers but rather focuses on speaker-defining information. To extract the log-f0 contour from the target speech recordings, we used the RAPT algorithm \cite{talkin1995robust} of the Speech Processing Toolkit (SPTK\footnote{\href{http://sp-tk.sourceforge.net}{http://sp-tk.sourceforge.net}}) with a threshold of 0 for voiced/unvoiced regions.

\subsection{Model training and fine-tuning}
\label{subsec:training_methodology}

Training a Voice Filter model that can generate speech from 1 minute of unseen speaker data is a two-step process: (1) background model training (Figure~\ref{fig:vf_training_pipeline}, Pre-train VF) and (2) fine-tuning to the minute of target unseen speaker data (Figure~\ref{fig:vf_training_pipeline}, Fine-tune VF).
\pagebreak

\begin{enumerate}
	\item The background Voice Filter, is trained in a one-to-many fashion for 1 million steps on the entire synthetic parallel multi-speaker corpus generated in the previous step. This model is capable of converting to any of the speakers seen during training but isn't robust enough to generalize to unseen speakers without further processing.
	\item We adapt the background model to the target Voice Filter by fine-tuning all the parameters of the background Voice Filter for 1000 steps on the target speaker's single minute of speech in a one-to-one fashion. We use the centroid of the target utterance-level speaker embeddings as we observed that, in our few-shot scenario, fine-tuning on a constant speaker embedding rather than on variable utterance-level embeddings resulted in more stable models. We have not tested the impact in quality on non-target speakers after fine-tuning, but we consider the resulting target Voice Filter to be speaker-dependent.
 \end{enumerate}
 
Both the background and target Voice Filter models are trained using the L1 spectral loss and the ADAM optimiser with default settings.

\subsection{Model Inference}
\label{subsec:inference_methodology}
\begin{figure}[!t]
\begin{center}
\includegraphics[scale=0.12]{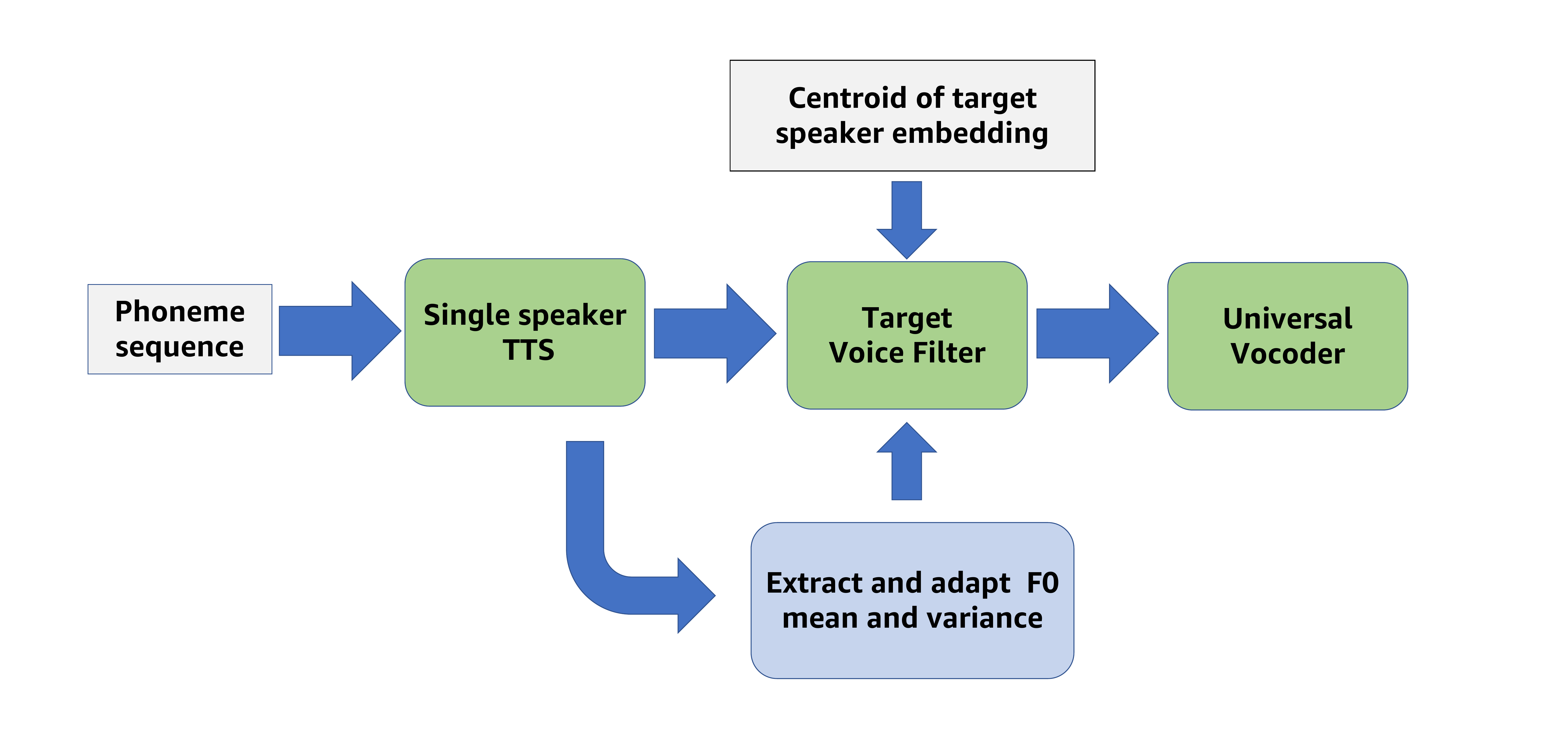}
\caption{Proposed Voice Filter inference flowchart.}
\label{fig:vf_inference_flowchart}
\end{center}
\end{figure}

Inference on the complete model (Figure~\ref{fig:vf_inference_flowchart}) requires us to run several models in succession: 
\begin{enumerate}
	\item Inference of the source Mel-spectrogram for the desired text and predicted durations with the single-speaker TTS model.
	\item Estimation of f0 from the source Mel-spectrogram and re-normalization to the mean and variance of the target speaker.
	\item Conversion of the source Mel-spectrogram to the target speaker by the fine-tuned Voice Filter.
	\item Synthesis of the voice-converted Mel-spectrogram to time-domain waveform via the vocoder.
\end{enumerate}

At this moment, we do not adapt speaking rate or phone durations to those of the target speaker as they are not simple to estimate on extremely low-resource scenarios and lead to significant artefacts.

\section{Experimental setup}

Models were assessed using both objective and perceptual metrics. For our evaluations we have used 4 male and 4 female speakers with 50 test utterances per speaker, resulting in 400 prompts overall.

Signal quality is objectively measured using the conditional Fréchet Speech Distance (cFSD) \cite{binkowski2019high}.
Specifially, a pre-trained XLSR-53 \cite{ott2019fairseq} wav2vec-2.0 \cite{baevski2020wav2vec} model is used to generate the activation distributions for recordings and synthesised samples. The distributions are then compared with a Fréchet Distance, which provides a measure of how close the generated speech is to an actual recording.
To objectively estimate a speaker similarity metric, we used the mean cosine distance between speaker embeddings (CSED) of recordings and predicted samples.
\begin{sloppypar}
MUltiple Stimuli with Hidden Reference and Anchor (MUSHRA) tests were used to perceptually assess naturalness, signal quality, speaker and speaking style similarity.
Samples from the systems being evaluated were presented to participants side by side.
They were asked to score them on a scale from 0 (the worst) to 100 (the best) in terms of the metric being evaluated.
We used the crowd-sourcing platform ClickWorker to assess each test utterance by a panel of 25 listeners.
Target speaker recordings were always included as a hidden upper-anchor system and we did not enforce requirements for at least one system to be rated 100. We provided the listeners with a reference sample for both the speaker and style similarity evaluations. The lower-anchor for the speaker similarity evaluation was the voice-converted samples of the furthest same gender speaker in the speaker embedding space. The lower-anchor for the style similarity evaluation was the un-filtered TTS system.
Paired two-sided Student T-tests with Holm-Bonferroni correction were used to validate the statistical significance of the differences between two systems at a p-value threshold of 0.05.
\end{sloppypar}

\section{Results}

\subsection{Extreme-low resource speech synthesis performance}\label{vs_others}

We compared our system with two multi-speaker state-of-the-art technologies that have shown high performance for low-resource speech synthesis: CopyCat (CC) \cite{karlapati2020copycat} model with an additional f0 conditioning\footnote{Preliminary results indicated that f0-conditioned CopyCat showed better signal quality stability and constituted a fairer comparison.} and a duration-controllable multi-speaker TTS (MS-TTS) \cite{shah21_ssw} model without the data augmentation component.
Both models were trained on the same dataset and conditions described in section \ref{subsec:training_methodology}, including fine-tuning for 1000 steps on the 1 minute of target speaker data.

Objective metrics and perceptual MUSHRA evaluation scores are reported in tables \ref{tab:objective_vs_dr2021} (rows 1-3) and \ref{tab:subjective_other}, respectively.
We observe a statistical preference in the MUSHRA evaluation for the proposed system in terms of all evaluated metrics. The objective metrics are aligned with these observations, indicating that our proposed method outperforms other speaker adaptation techniques when using the same amount of data.

\begin{table}[h!]
	\centering
	\resizebox{0.90\columnwidth}{!}{%
		\begin{tabular}{lcc}
			\toprule
			\textbf{System} & \textbf{Speaker sim. (CSED)} & \textbf{Signal quality (cFSD)} \\ \midrule
			\textbf{VF}     & $\mathbf{0.192}$                     & $\mathbf{0.197}$                 \\
			CC              & $0.198$                              & $0.249         $                 \\
			MS-TTS          & $0.207$                              & $0.263         $             \\ 
			TTS-DA          & $0.205$                              & $0.224         $                 \\ \bottomrule
		\end{tabular}%
	}
	\caption{Average objective metrics for all evaluated systems. Best numbers are highlighted in bold. TTS-DA was trained on 30 minutes of speech instead of 1.}
	\label{tab:objective_vs_dr2021}
\end{table}

\begin{table}[h!]
\centering
\resizebox{0.90\columnwidth}{!}{%
\begin{tabular}{lcccc}
\toprule
\textbf{System} & \textbf{Sp. sim.} & \textbf{Style sim.} & \textbf{Nat.}   & \textbf{Sig. Q.}  \\ \midrule
Rec             & $79.43 \pm 0.66$             & $75.53 \pm 0.73$               & $78.72 \pm 0.64$           & $76.67 \pm 0.75$            \\
\textbf{VF}              & $\mathbf{67.96} \pm 0.86$    & $\mathbf{64.4} \pm 0.82$       & $\mathbf{53.09} \pm 0.81$  & $\mathbf{56.28} \pm 0.84$   \\
CC              & $66.57 \pm 0.88$             & $63.56 \pm 0.84$               & $52.08 \pm 0.81$           & $55.28 \pm 0.85$            \\
MS-TTS          & $65.83 \pm 0.91$             & $62.10 \pm 0.86 $              & $50.71 \pm 0.81$           & $54.24 \pm 0.85$            \\
Lower-anchor    & $37.90 \pm 1.00$             & $37.96 \pm 1.12$               & $-$                        & $-$                 \\ \bottomrule
\end{tabular}%
}
\caption{Average MUSHRA results with confidence interval of 95\%. Best scores with statistical difference between Voice filter (VF) and reference systems are highlighted in bold ($\textit{p} < 0.05$).}
\label{tab:subjective_other}
\end{table}

\subsection{Ablation study on data quantity}\label{sec:impact_of_the_data}

In order to understand the impact of the extremely low-resource scenario, we trained the Voice Filter using 1, 5 and 25 minutes of target speaker data during fine-tuning. Objective metrics and perceptual MUSHRA evaluation scores are reported in tables \ref{tab:objective_varying} and \ref{tab:subjective_varying}, respectively.

In the subjective evaluation, listeners did not perceive a statistically significant difference between the different data scenarios, although objective metrics still show slight improvements with bigger amounts of data. This hints that, while there may be room to improve the performance of the system, the nature of our Voice Filter doesn't benefit perceptually from richer data scenarios.
With only 1 minute of target data, we are able to create high-quality samples.

\begin{table}[h!]
\centering
\resizebox{0.90\columnwidth}{!}{%
\begin{tabular}{lcccc}
\toprule
\textbf{\# min} & \textbf{Speaker sim. (CSED)} & \textbf{Signal quality (cFSD)} \\ \midrule
1 min           & $0.192$                        & $0.197         $                \\
5 min           & $\mathbf{0.183}$               & $0.189         $                 \\
25 min          & $0.185$                        & $\mathbf{0.176}$               \\  \bottomrule
\end{tabular}%
}
\caption{Average objective metrics for Voice Filter trained using varying amount of data. Best results are highlighted in bold.}
\label{tab:objective_varying}
\end{table}

\begin{table}[h!]
\centering
\resizebox{0.90\columnwidth}{!}{%
\begin{tabular}{lcccc}
\toprule
\textbf{\# min} & \textbf{Sp. sim.} & \textbf{Style sim.} & \textbf{Nat.} & \textbf{Sig. Q.} \\ \midrule
Rec             & $73.77 \pm 0.99$    & $72.06 \pm 0.95$              &  $78.69 \pm 0.70$       & $73.94 \pm 0.91$                \\
1 min           & $51.90 \pm 0.99$    & $54.67 \pm 0.98$              &  $55.10 \pm 0.84$       & $53.94 \pm 0.95$                \\
5 min           & $51.93 \pm 1.01$    & $55.36 \pm 0.97$              &  $55.06 \pm 0.83$       & $53.83 \pm 0.97$                \\
25 min          & $52.00 \pm 1.01$    & $54.75 \pm 0.99$              &  $55.21 \pm 0.82$       & $53.63 \pm 0.95$                \\ \bottomrule
\end{tabular}%
}
\caption{Average MUSHRA results with confidence interval of 95\% for Voice Filter trained with varying amount of data. No statistically significant differences ($\textit{p} < 0.05$) between systems.}
\label{tab:subjective_varying}
\end{table}

\subsection{Comparison against a competitive TTS}\label{vs_dr}

Finally, we investigate the quality of the generated speech when compared to a TTS system trained on a larger amount of target recordings.
For that, we compared Voice Filter against a validated low-resource TTS technology that showed to be competitive when trained on 30 minutes of target speech (TTS-DA) \cite{huybrechts2021low}.
It is worth noting that such a technology implicitly results in the TTS system estimating the phone duration for the target speaker, which is not the case for the proposed Voice Filter.
Effectively this means we are comparing the proposed Voice Filter against a system trained on 30 times more target data that has been validated also to be competitive against TTS voices trained on 5+ hours of target recordings.

Objective metrics and perceptual MUSHRA evaluation scores are reported in tables \ref{tab:objective_vs_dr2021} (rows 1 \& 4) and \ref{tab:subjective_vs_dr2021}, respectively.
While we observe a 4\% relative  degradation in speaker similarity, the MUSHRA evaluations indicate no statistical difference between the systems in terms of signal quality, naturalness, and style similarity. On the other hand, in terms of objective metrics, we observe that speaker similarity and signal quality are better for our proposed method.
Overall, results show that our model is on par with TTS-DA with a slight human-perceived degradation in terms of speaker similarity, and this despite the much smaller target training dataset being used.

\begin{table}[h!]
	\centering
	\resizebox{0.90\columnwidth}{!}{%
		\begin{tabular}{lcccc}
			\toprule
			\textbf{System} & \textbf{Sp. sim.}       & \textbf{Style sim.}     & \textbf{Nat.}           & \textbf{Sig. Q.} \\ \midrule
			Rec             & $83.23           \pm 0.66$  & $79.82 \pm 0.80$             & $77.05 \pm 0.56$        & $76.75 \pm 0.52$          \\
			VF              & $70.38           \pm 0.93$  & $69.53 \pm 0.98$             & $55.24 \pm 0.80$        & $55.60 \pm 0.72$          \\
			TTS-DA          & $\mathbf{73.67}  \pm 0.85$  & $69.97 \pm 1.01$             & $55.21 \pm 0.81$        & $55.83 \pm 0.71$          \\
			Lower-anchor    & $37.74           \pm 1.10$  & $39.51 \pm 1.46$             & $-$                     & $-$                       \\ \bottomrule
		\end{tabular}%
	}
	\caption{Average MUSHRA results. Best scores with statistical difference between Voice filter (VF) and reference systems are highlighted in bold ($\textit{p} < 0.05$)}
	\label{tab:subjective_vs_dr2021}
\end{table}
\section{Conclusions}

In this work, we proposed a novel extremely low-resource TTS method called Voice Filter that can produce high-quality speech when using only 1 minute of speech.
 
Voice Filter splits the TTS process in a speech content and speaker identity generation task. The speaker identity is generated via a fine-tuned one-to-many VC module, which makes it easily scalable to new speakers even on extremely low-resource settings. 
The speech content generation module is a duration-controllable single-speaker TTS system that has the added benefit of enabling us to generate a synthetic parallel corpus. This enables Voice Filter to work in a parallel frame-level condition, which has a higher quality ceiling and lower modelling complexity.
Evaluations show that our Voice Filter outperforms other few-shot speech synthesis techniques in terms of objective and subjective metrics on the 1-minute data scenario, with quality comparable to a SOTA system trained on 30 times more data.

In conclusion, we consider the Voice Filter paradigm to be a first step towards building extremely low-resource TTS as a post-processing VC plug-in. Moreover, we believe that the generation of synthetic parallel duration-controllable data will enable further scenarios in speech technologies that were limited previously because of data availability.

\bibliographystyle{IEEEbib}
\bibliography{mybib.bib}

\end{document}